\documentclass
[prl,twocolumn,10pt,tightenlines,showpacs,letterpaper,runinaddress]{revtex4}
\usepackage{graphicx,amsfonts,amssymb}
\usepackage{amsmath}
\usepackage{latexsym}

\def\b{\begin{eqnarray}}
\def\e{\end{eqnarray}}
\def\n{\noindent}
\begin{document}
\title[Two-component Camassa-Holm equation]{On an integrable two-component Camassa-Holm shallow water system}
\author{Adrian Constantin $^{1,2}$ and Rossen I. Ivanov $^{3,2}$}
\affiliation{$^{1}$Faculty of Mathematics, University of Vienna,
Nordbergstrasse 15, 1090 Vienna, Austria, \\
email: adrian.constantin@univie.ac.at\\
\\
$^{2}$Department of Mathematics, Lund University,
22100 Lund, Sweden,\\
\\
$^{3}$School of Mathematical Sciences, Dublin Institute of
Technology, Kevin Street, Dublin 8, Ireland,
\\
e-mail: rivanov@dit.ie} 

\begin{abstract}
The interest in the Camassa-Holm equation inspired the search for
various generalizations of this equation with interesting
properties and applications. In this letter we deal with such a
two-component integrable system of coupled equations. First we
derive the system in the context of shallow water theory. Then we
show that while small initial data develop into global solutions,
for some initial data wave breaking occurs. We also discuss the
solitary wave solutions. Finally, we present an explicit
construction for the peakon solutions in the short wave limit of
system.

\end{abstract}
\maketitle

\n In recent years the Camassa-Holm (CH) equation \cite{CH93}
\begin{equation} m_t+\omega u_{x}+2mu_x+m_xu=0,\quad m=u-u_{xx},
\label {CH_1}
\end{equation}
($\omega$ being an arbitrary constant) has caught a great deal of
attention. It is a nonlinear dispersive wave equation that models
the propagation of unidirectional irrotational shallow water waves
over a flat bed \cite{CH93,DGH01,J02,CL}, as well as water waves
moving over an underlying shear flow \cite{J03}. The CH equation
also arises in the study of a certain non-Newtonian fluids
\cite{B99} and also models finite length, small amplitude radial
deformation waves in cylindrical hyperelastic rods \cite{Dai98}.
The CH equation has a bi-Hamiltonian structure \cite{FF80} (and an
infinite number of conservation laws), it is completely integrable
(see \cite{CH93} for the Lax pair formulation and \cite{CGI} for
the direct/inverse scattering approach), and its solitary wave
solutions are solitons \cite{CH93, BSS99, CS, CS1, Pa} with stable
profiles \cite{CS, CS1}. The equation attracted a lot of attention
in recent years due to two remarkable features. The first is the
presence of solutions in the form of peaked solitary waves or
'peakons' \cite{CH93,BSS99,L} for $\omega=0$: the peakon
$u(x,t)=ce^{-|x-ct|}$ travelling at finite speed $c\ne 0$ is
smooth except at its crest, where it is continuous, but has a jump
discontinuity in its first derivative. The peakons replicate a
characteristic of the travelling waves of greatest height - exact
travelling solutions of the governing equations for water waves
with a peak at their crest \cite{T,C1,CE1} whose capture by
simpler approximate shallow water models has eluded researchers
until recently \cite{W}. A further remarkable property of the CH
equation is the presence of breaking waves (i.e. the equation has
smooth solutions which develop singularities in finite time in the
form of breaking waves \cite{CH93,CE,M} - the solution remains
bounded while its slope becomes unbounded in finite time \cite{W})
as well as that of smooth solutions defined for all times
\cite{C}. These two phenomena have always fascinated the fluid
mechanics community: 'Although breaking and peaking, as well as
criteria for the occurrence of each, are without doubt contained
in the equation of the exact potential theory, it is intriguing to
know what kind of simpler mathematical equation could include all
these phenomena' \cite{W}. The short wave limit of the CH equation
is the Hunter-Saxton (HS) equation \b
u_{xxt}+2u_xu_{xx}+uu_{xxx}=0, \label{HS}  \e obtained from
(\ref{CH_1}) by taking $m=-u_{xx}$. It describes the propagation
of waves in a massive director field of a nematic liquid crystal
\cite{HS}, with the orientation of the molecules described by the
field of unit vectors $(\cos u(x,t), \sin u(x,t))$, where $x$ is
the space variable in a reference frame moving with the linearized
wave velocity, and $t$ is a 'slow time variable'.

The equations (\ref{CH_1}), (\ref{HS}) admit many integrable
multicomponent generalizations \cite{I06}, the most popular of
which is \b  m_{t}+2u_{x}m+um_{x}+\sigma\rho \rho_x=0, \label{1}\\
\rho_{t}+(u\rho)_{x}=0, \label{2} \e where $ m=\sigma_1 u-u_{xx}$,
$\sigma=\pm 1$ and $\sigma_1=1$ or, in the 'short wave' limit,
$\sigma_1=0$. (The CH equation can be obtained via the obvious
reduction $\rho\equiv 0$.) This system appears originally in
\cite{OR} and its mathematical properties have been studied
further in many works, e.g. \cite{SA,CLZ05,I06,ELY,Ar}. The system
is integrable -- it can be written as a compatibility condition of
two linear systems (Lax pair) with a spectral parameter $\zeta$:
\b \Psi_{xx}=\Big(-\sigma\zeta^2\rho^2+\zeta m
+\frac{\sigma_1}{4}\Big)\Psi, \nonumber \\
\Psi_{t}=\Big(\frac{1}{2\zeta}-u\Big)\Psi_x+\frac{1}{2}u_x\Psi.
\nonumber  \e It is bi-Hamiltonian, the first
Poisson bracket \b \{F_1,F_2\}=\phantom{****************************} \nonumber \\
\!\!-\!\!\int\Big[\frac{\delta F_1}{\delta m}(m\partial+\partial
m)\frac{\delta F_2}{\delta m}+\frac{\delta F_1}{\delta m}\rho
\partial\frac{\delta F_2}{\delta \rho}+\frac{\delta F_1}{\delta \rho}\partial \rho \frac{\delta F_2}{\delta
m}\Big]\text{d}x, \nonumber \e corresponding to the Hamiltonian \b
H=\frac{1}{2}\int (um+\sigma \rho^2)\text{d}x \nonumber \e and the
second Poisson bracket \b \{F_1,F_2\}_2=-\int\Big[\frac{\delta
F_1}{\delta m}(\partial-\partial^{3})\frac{\delta F_2}{\delta
m}+\frac{\delta F_1}{\delta \rho}
\partial\frac{\delta F_2}{\delta \rho}\Big]\text{d}x  \nonumber \e
corresponding to the Hamiltonian \b H_2=\frac{1}{2}\int (\sigma
u\rho^2+u^3+uu_x^2)\text{dx}. \nonumber \e There are two Casimirs:
$\int \rho \text{d}x$ and $\int m \text{d}x$.

In what follows, we are going to demonstrate how the system
(\ref{1}), (\ref{2}) arises in shallow water theory. We start from
the Green-Naghdi (GN) equations \cite{GN76, J02}, which are
derived from the Euler's equations under certain assumptions, as
follows. Consider the motion of shallow water over a flat surface,
which is located at $z=0$ with respect to the usual Cartesian
reference frame. We assume that the motion is in the $x$-direction
and the physical variables do not depend on $y$. Let $h$ be the
mean level of water, $a$ -- the typical amplitude of the wave and
$\lambda$ -- the typical wavelength of the wave. Let us now
introduce the dimensionless parameters $\varepsilon=a/h$ and
$\delta=h/\lambda$, which are supposed to be small in the shallow
water regime. The variable $u(x,t)$ describes the horizontal
velocity of the fluid, $\eta(x,t)$ describes the horizontal
deviation of the surface from equilibrium, all measured in
dimensionless units. The GN equations \b u_t\!\!&+&\!\!\varepsilon
u u_x +\eta_x=\frac{\delta^2/3}{1\!\!+\!\!\varepsilon
\eta}[(1+\varepsilon \eta)^3(u_{xt}+\varepsilon u
u_{xx}-\varepsilon u_x^2)]_x,
\nonumber \\
\eta_{t}\!\!&+&\!\![(u(1+\varepsilon \eta)]_{x}=0. \nonumber \e
are obtained under the assumption that at leading order $u$ is not
a function of $z$. The leading order expansion with respect to the
parameters $\varepsilon$ and $\delta^2$ gives the system
\b \Big(u-\frac{\delta^2}{3}u_{xx}\Big)_{t}+\varepsilon u u_{x}+\eta_x=0, \label{3}\\
\eta_{t}+[(u(1+\varepsilon \eta)]_{x}=0. \label{4} \e  One can
demonstrate that the system (\ref{3}), (\ref{4}) can be related to
the system (\ref{1}), (\ref{2}) in the first order with respect to
$\varepsilon$ and $\delta^2$. Indeed, let us define \b
\rho=1+\frac{1}{2}\varepsilon
\eta-\frac{1}{8}\varepsilon^2(u^2+\eta^2).\nonumber \e The
expansion of $\rho^2$ in the same order of $\varepsilon$ is \b
\rho^2=1+\varepsilon \eta-\frac{1}{4}\varepsilon^2 u^2.\nonumber
\e With this definition it is straightforward to write (\ref{3})
in the form \b
\Big(u-\frac{\delta^2}{3}u_{xx}\Big)_{t}+\frac{3}{2}\varepsilon u
u_{x}+\frac{1}{\varepsilon}(\rho^2)_x=0,\nonumber \e or,
introducing the variable $m=u-\frac{1}{3}\delta^2 u_{xx}$, at the
same order (i.e. neglecting terms of order $\varepsilon \delta^2$)
\b m_{t}+\varepsilon m u_x+\frac{1}{2}\varepsilon u
m_{x}+\frac{1}{\varepsilon}(\rho^2)_x=0.\label{5} \e Next, using
the fact that $u_t\approx -\eta_x$, $\eta_t\approx -u_x$, from the
definition of $\rho$ we get $\rho_t=\frac{1}{2}\varepsilon \eta_t
+\frac{1}{4}\varepsilon^2(\eta u)_x$. With this expression for
$\rho_t$ and with $\rho \approx 1+\varepsilon u$, equation
(\ref{4}) can be written as \b \rho_{t}+\frac{\varepsilon}{2}(\rho
u)_x=0.\label{6} \e The rescaling $u\rightarrow
\frac{2}{\varepsilon} u$, $x\rightarrow \frac{\delta}{\sqrt{3}}x$,
$t\rightarrow \frac{\delta}{\sqrt{3}} t$ in (\ref{5}), (\ref{6})
gives (\ref{1}), (\ref{2}) with $\sigma=\sigma_1=1$. The case
$\sigma=-1$, which is often considered, corresponds to the
situation in which the gravity acceleration points upwards. We
mention also that the Kaup - Boussinesq system \cite{K76} is
another integrable system matching the GN equation to the same
order of the parameters $\varepsilon, \delta$ \cite{W,EGP}. Notice
that in the hydrodynamical derivation of (3)-(4) we require that
$u(x,t) \to 0$ and $\rho(x,t) \to 1$ as $|x| \to \infty$, at any
instant $t$.

We will now show that for the system (\ref{1}), (\ref{2}) in the
hydrodynamically relevant case  $\sigma=\sigma_1=1$ wave breaking
is the only way that singularities arise in smooth solutions. The
system admits breaking wave solutions as well as solutions defined
for all times. In particular, we will analyze the traveling wave
solutions

The well-posedness (existence, uniqueness, and continuous
dependence on data) follows by Kato's semigroup theory \cite{K}
for initial data $u_0=u(\cdot,0) \in H^3$ and
$\rho_0=\rho(\cdot,0)$ such that $(\rho_0-1) \in H^2$
\cite{foot2}. If $T=T(u_0,\rho_0)>0$ is the maximal existence
time, then the integral of motion
\begin{equation}\label{im}
\int[u^2+u_x^2+(\rho-1)^2]\text{d}x
\end{equation}
ensures that $u(\cdot,t)$ is uniformly bounded (i.e. for all
values of $x \in \mathbb{R}$ and all $0\le t<T$) in view of the
inequality
\begin{equation}\label{i}
\sup_{x \in \mathbb{R}}\,|u(t,x)|^2 \le \frac{1}{2}\,\int
(u^2+u_x^2)\text{d}x.
\end{equation}
Considerations analogous to those made in \cite{ELY} for a similar
system show that the solution blows up in finite time (i.e.
$T<\infty$) if and only if
\begin{equation}\label{b1}
\liminf_{t \uparrow T} \,\{u_x(t,x)\}=-\infty,
\end{equation}
which, in light of the uniform boundedness of $u$, is interpreted
as wave breaking.

To show that wave breaking occurs, we introduce the family
$\{\varphi(\cdot,t)\}_{t \in [0,T)}$ of diffeomorphisms
$\varphi(\cdot,t): \mathbb{R} \to \mathbb{R}$ defined by
\begin{equation}\label{d}
\partial_t \,\varphi(x,t)=u(\varphi(x,t),t),\quad \varphi(x,0)=x,
\end{equation}
and we denote
$$M(x,t)=u_x(\varphi(x,t),t),\quad \gamma(x,t)=\rho(\varphi(x,t),t).$$
Consider now initial data satisfying $\rho_0(0)=0$ and
\begin{equation}\label{c}
u_0'(0)<-2\,\Big( || u_0||_1^2+||\rho-1||_0^2\Big)^{1/2}.
\end{equation}
Noticing that $(1-\partial_x^2)^{-1}f=p \ast f$ (convolution) with
$p(x)=\displaystyle\frac{1}{2}\,e^{-|x|}$, and applying the
operator $(1-\partial_x^2)^{-1}$ to (3), we get \b u_t+uu_x + p
\ast (u^2+\frac{1}{2}\,u_x^2+\frac{1}{2}\,\rho^2)=0.\nonumber \e
Applying now $\partial_x$ and using the identity $\partial_x^2 \,p
\ast f=p \ast f -f$, we obtain
\begin{equation}\label{m}
u_{tx}\!+\!uu_{xx}\!+\!\frac{1}{2}u_x^2=\frac{1}{2}(u^2\!+\!\rho^2)-p
\ast (u^2\!+\!\frac{1}{2}u_x^2\!+\!\frac{1}{2}\rho^2).
\end{equation}
This equation in combination with (\ref{d}) yields
\begin{equation}\label{i2}
\partial_t M(t,x)\!+\!\frac{1}{2}M^2(t,x) \!\le
\!\frac{1}{2}\Big(u^2(\varphi(x,t),t)\!+\!\gamma^2(x,t)\Big).
\end{equation}
On the other hand, from (\ref{d}) and (4), we obtain
\begin{equation}\label{t}
\partial_t \gamma=-\gamma M.
\end{equation}
Since $\gamma(0,0)=0$ we infer that $\gamma(0,t)=0$ for $0\le t <
T$. The relation (\ref{c}) together with (\ref{im}), (\ref{i})
ensure that $4\,u^2(\varphi(x,t),t) \le M^2(0,0)$. But then
(\ref{i2}) yields \b \partial_t \,M(0,t) \le
-\,\frac{1}{4}M^2(t,0)\qquad \text{for} \qquad 0\le t < T.
\nonumber \e As $M(0,0)=u_0'(0)<0$, this implies \b M(0,t) \le
\displaystyle\frac{4u_0'(0)}{4+u_0'(0)\,t} \to -\infty \nonumber
\e in finite time.

However, not all solutions develop singularities in finite time.
For example, if the initial data is sufficiently small, then the
solution evolving from it is defined for all times. More
precisely, let $\alpha \in (0,1)$ and assume that $|1-\rho_0(x)|
\le \alpha$ for all $x \in \mathbb{R}$, while $||u_0||_2
+||\rho_0-1||_0 \le \alpha$. Then the corresponding solution to
(\ref{1}), (\ref{2}) is global in time. Indeed, if the maximal
existence time were $T<\infty$, then for some $x_0 \in \mathbb{R}$
we would have $\liminf_{t \uparrow T}\,M(t,x_0)=-\infty$. We now
show that this is impossible. First, notice that
\begin{equation}\label{bu}
u^2(x,t) \le \alpha^2,\quad (x,t) \in \mathbb{R} \times [0,T).
\end{equation}
by (\ref{im}), (\ref{i}). The integral of motion (\ref{im}) also
ensures that $||\rho(\cdot,t)-1||_0 \le \alpha$ on $[0,T)$. Using
this and (\ref{2}), (\ref{im}), (\ref{bu}), we get  \b &0& \le p
\ast (u^2+\frac{1}{2}\,[u_x^2+\rho^2]) \nonumber \\ &\le&
\frac{1}{2} p \ast u^2 + \frac{1}{2}(u^2+u_x^2+(\rho\!\!-\!\!1)^2)
+ p \ast (\rho\!\! -\!\!1) + \frac{1}{2}\,p \ast 1 \nonumber \\
&\le& \alpha^2 + \alpha +1\nonumber \e on $\mathbb{R} \times
[0,T)$. We now infer from (\ref{d}), (\ref{m}) that at $x=x_0$,
\begin{equation}\label{f}
M_t = -\frac{1}{2}\,M^2 + \frac{1}{2}\,\gamma^2 -f(t),\quad t \in
[0,T),
\end{equation}
with the continuous function $f: [0,T) \to [0,\infty)$ bounded,
i.e. there is a constant $k_0>0$ such that $0 \le \gamma(t) \le
k_0$ on $[0,T)$. The solution $\Big(\gamma(t),\,M(t)\Big)$ of the
nonlinear system (\ref{t}), (\ref{f}) with initial data
$\gamma(0)=\rho_0(x_0)>0$ and $M(0)=u_0'(x_0)$ supposedly blows up
in finite time as $\liminf_{t \uparrow T}\,M(t)=-\infty$. However,
notice that (\ref{t}) and $\gamma(0)>0$ ensure $\gamma(t) >0$ for
all $t \in [0,T)$. Thus we may consider the positive function \b
w(t)=\gamma(0)\gamma(t) +
\frac{\gamma(0)}{\gamma(t)}[1+M^2(t)]\nonumber \e for $t \in
[0,T)$. Using (\ref{t}), (\ref{f}) we get \b
w'(t)&=&2\,\frac{\gamma(0)}{\gamma(t)}M(t)[f(t)+\frac{1}{2}]\nonumber
\\ & \le& [k_0+1]\frac{\gamma(0)}{\gamma(t)}(1+M^2)\le
[k_0+1]w(t)\nonumber \e for all $t \in (0,T)$. Thus \b w(t) \le
w(0)\,\exp( [k_0+1] t)\nonumber \e on $[0,T)$ and this prevents
blowup. The obtained contradiction shows that the solution is
defined globally in time.

Global existence is however not confined to small amplitude waves,
as we shall see now by establishing the existence of traveling
waves of large amplitude: solutions
$u(x,t)=\psi(x-ct),\,\rho(x,t)=\xi(x-ct)$ traveling with constant
wave speed $c>0$. To find whether such solutions exist, notice
that with the previous Ansatz equation (4) becomes
$\xi'(c-\psi)=\xi\psi'$ and the asymptotic limits $\psi(x) \to 0$
and $\xi(x) \to 1$ as $|x| \to \infty$ yield \b
\xi=\frac{c}{c-\psi}.\nonumber \e Thus (3) becomes a differential
equation solely for the unknown $\psi$. Integrating this equation
on $(-\infty,x]$ and taking into account the asymptotic behaviour
of $\psi$, we get the equation \b
-c\psi+c\psi''+\frac{3}{2}\psi^2-\psi\psi''-\frac{1}{2}(\psi')^2+
\frac{c^2}{2(c-\psi)^2}=\frac{1}{2}.\nonumber \e Multiplication by
$\psi'$ and another integration on $(-\infty,x]$ leads to \b
\Big((\psi')^2-\psi^2\Big)\,(c-\psi)+\frac{c^2}{c-\psi}=\psi+c,\nonumber
\e recalling the decay of $\psi$ far out. Thus
\begin{equation}\label{ode}
(\psi')^2=\frac{\psi^2}{(c-\psi)^2}\,(c-\psi-1)(c-\psi+1).
\end{equation}
The asymptotic behaviour $\psi(x) \to 0$ as $|x| \to \infty$
yields now the necessary condition $c \ge 1$ for the existence of
traveling waves. For $c \ge 1$, a qualitative analysis of
(\ref{ode}) shows that $0 \le \psi \le c-1$. Thus nontrivial
traveling waves exist only for $c>1$, in which case both $\psi$
and $\xi$ are smooth waves of elevation with a single crest
profile of maximal amplitude $c-1$, respectively $c$. It is
possible to find explicit formulas for the traveling waves in
terms of elliptic functions cf. \cite{hh}. Due to the
integrability of the system, we expect the solitary waves to
interact like solitons.

Notice the absence of peakons among the solitary wave solutions.
However, there are peakon solutions of the 'short wave limit'
equation $\sigma_1=0$. Although this limit is not covered by the
presented hydrodynamical derivation, we will describe briefly the
construction of the peakon solutions, since these are interesting
by themselves. The limit $\sigma_1=0$ is a two component analog of
the Hunter-Saxton equation. Such system is a particular case of
the Gurevich-Zybin system \cite{GZ}, which describes the dynamics
in a model of nondissipative dark matter \cite{P05}.

The peakon solutions have the form
\b m(x,t)&=&\sum_{k=1}^N m_k(t) \delta(x-x_k(t)), \label{mk} \\
u(x,t)&=&-\frac{1}{2}\sum_{k=1}^N m_k(t) |x-x_k(t)|, \label{uk} \\
\rho(x,t)&=&\sum_{k=1}^N \rho_k(t) \theta(x-x_k(t)), \label{rhok}
\e where $\theta$ is the Heaviside unit step function. The
asymptotic behaviour $\rho(x,t)\rightarrow 0$ for
$x\rightarrow\infty$ and the condition $\int m \,\text{d}x=0$
(recall that $m=-u_{xx}$) lead to \b \sum_{l=1}^N m_l=\sum_{l=1}^N
\rho_l=0, \nonumber \e or \b \sum_{l=1}^N \mu_l=0 \nonumber \e in
terms of the new complex variable $\mu_k\equiv m_k+i\rho_k$. The
substitution of the Ansatz (\ref{mk}) -- (\ref{rhok}) into
(\ref{1}), (\ref{2}), under the assumption that
$x_1(t)<x_2(t)<\ldots <x_N(t)$ for all $t$, (a condition holding
for the peakons of (\ref{HS}) cf. \cite{BSS01}) gives the
following dynamical system for the time-dependent variables: \b
\frac{\text{d}x_k}{\text{d}t}&=&-\frac{1}{2}\sum_{l=1}^N m_l\,
|x_l-x_k|, \label{xdot} \\
\frac{\text{d}\mu_k}{\text{d}t}&=&\frac{\mu_k}{2}\sum_{l=1}^N\mu_l\,\text{sgn}(k\!-\!l)
, \label{mudot}\e with the convention $\text{sgn}(0)=0$. The
integrals for this system can be obtained from the integrals of
(\ref{1}), (\ref{2}) (available in \cite{I06}) by substituting the
expressions (\ref{mk}), (\ref{uk}), (\ref{rhok}). It is convenient
to write the system in terms of the new independent variables \b
\Delta_k\equiv x_{k+1}-x_k,\quad
M_k\equiv\mu_1+\ldots+\mu_k,\nonumber \e with $k=1,2,\ldots,N-1$.
The Hamiltonian of the new system is \b
H=\frac{1}{2}\sum_{l=1}^{N-1}|M_k|^2\Delta_k,\nonumber \e the
equations \b
\frac{\text{d}\Delta_k}{\text{d}t}=-\text{Re}(M_k)\,\Delta_k,
\qquad \frac{\text{d}M_k}{\text{d}t}= \frac{1}{2}\,M_k^2,
\nonumber \e are Hamiltonian with respect to the bracket \b
\{\Delta_k,M_l\}=-\frac{M_k}{\bar{M}_k}\delta_{lk}, \nonumber \e
in which the bar stands for complex conjugation. These equations
integrate immediately: \b
M_k(t) &=& -\frac{1}{t/2+c_k},\nonumber \\
\Delta_k(t) &=& \Delta_k(0)\,
\displaystyle\frac{(t/2+c_{k,1})^2+c_{k,2}^2}{c_{k,1}^2+c_{k,2}^2},
\nonumber \e where $c_k\equiv c_{k,1}+ic_{k,2}=-M_k^{-1}(0)$ is a
complex constant with real and imaginary parts $c_{k,1}$ and
$c_{k,2}$ respectively. Notice that the large time asymptotics
$$M_k \sim t^{-1},\quad\Delta_k\sim t^2,$$
are the same as those for the peakons of the Hunter-Saxton
equation (2) when $\rho_k\equiv 0$ (see \cite{BSS01}).

\subsection*{Acknowledgements}

The support of the G. Gustafsson Foundation for Research in
Natural Sciences and Medicine is gratefully acknowledged. R.I. is
thankful to Prof. R. Camassa, Prof. D.~D.~Holm and Dr.
G.~Grahovski for discussions.


\begin{thebibliography}{99}
\bibitem{foot2} For $n \ge 0$ integer $H^n$ is the Hilbert space of
square integrable functions $f: \mathbb{R} \to \mathbb{R}$ with
square integrable distributional derivatives up to order $n$,
endowed with the norm $||f||_n=\int \sum_{j=0}^n (\partial_x^j
f)^2(x)\,dx\Big)$. If $f \in H^n$ with $n \ge 1$, then $f$ has
continuous derivatives up to order $(n-1)$ and $f$ decays to zero
at infinity.
\end{thebibliography}
\end{document}